# Probing Thermally Activated Atomic and Nanocrystalline Defect Motion through Noise Processes in RuO$_2$ Nanowires

Sheng-Shiuan Yeh,[1,2,3] Cheng-Ya Yu,[4] Yi-Te Lee,[1] Shao-Pin Chiu,[1] Juhn-Jong Lin[1,2,4,*]

[1]Institute of Physics, National Chiao Tung University, Hsinchu 30010, Taiwan

[2]Center for Emergent Functional Matter Science, National Chiao Tung University, Hsinchu 30010, Taiwan

[3]International College of Semiconductor Technology, National Chiao Tung University, Hsinchu 30010, Taiwan

[4]Department of Electrophysics, National Chiao Tung University, Hsinchu 30010, Taiwan

[*]Correspondence to: jjlin@mail.nctu.edu.tw

## ABSTRACT

The present-day nanodevice dimensions continuously shrink, with the aim to prolong Moore's law. As downsizing meticulously persists, undesirable dynamic defects, which cause low-frequency noise and structural instability, play detrimental roles on limiting the ultimate performance and reliability of miniaturized devices. A good understanding and a meaningful control of the defect kinetics then become fundamental and urgent issues. Here we report observations of thermally activated atomic defect motion as well as nanocrystalline defect motion through electrical noise processes in metallic RuO$_2$ rutile nanowires around room temperature. First, we extract the energy distribution function and the number density of mobile atomic defects (oxygen vacancies). Second, we obtain the geometrical size, grain-boundary bonding strength, and relaxation times of dynamic nanocrystallites. Our results show clearly a powerful probe for effective and noninvasive characterizations of nanostructures and nanomaterials for which quantitative information about mechanical hardness, breakdown current density, and/or resistance noise is essential.

# I. INTRODUCTION

Metallic nanowires (NWs) are prime candidates for integrating with nanomechanical resonators to implement applications as, among others, ultrasensitive force sensors [1], electron-spin flip detectors [2], and single molecule detectors [3]. They also serve as indispensable advanced microelectronics interconnections [4-6]. For every application, the undesirable dynamical structural defects (or, dynamic defects for short) contained in the NWs play detrimental roles on limiting the device performance. For example, dynamic defects will cause energy dissipation and hamper the quality factor of a nanomechanical resonator, thus reducing its resolution power [7]. Defect motion also leads to resonance frequency instability [8]. In electrical conductors, small and large mobile defects generate resistance fluctuations, causing low-frequency ($1/f$) noise or random telegraph noise (RTN), where $f$ denotes frequency [9]. Both $1/f$ noise and RTN are deemed to deteriorate the reliability and quality of a nanoelectronic device [10].

In real materials, dynamic defects can be single atoms or clusters of atoms, which act as moving scattering centers. They are validly modeled as two-level systems (TLSs) [11,12]. A TLS is a double-well potential having a barrier height $E_i$ ($i$ = 1, 2), see Fig. 1(a). The mobile object spontaneously and repeatedly switches between the two wells, with characteristic fluctuating times $\tau_i$. In this work, we shall call a mobile atomic defect an "atomic TLS," and a mobile cluster of atoms a "granular TLS." In a given device, there usually exist a huge amount of dynamic atomic defects. These atomic TLSs have a wide range of $E_i$ and $\tau_i$ values, which are described by an energy distribution function $g(E)$, where $g(E)$ is defined as the number of atomic TLSs per unit energy $E$ per unit volume. On the other hand, while mobile clusters of atoms may also exist in a given device, they are usually much more difficult to detect, partly because $\tau_i$ may be relatively long and the switching events are comparatively rare (within the experimental bandwidth). Thus far, quantitative reports on the $g(E)$ function for atomic TLSs and the defect kinetics of granular TLSs in metallic NWs have been barely available. Previously, atomic TLSs were observed in, *e.g.*, Bi [13], Al [14] and Cu [15,16] films, AuPd NWs [17], as well as Cu nanoconstrictions [18]. Granular TLSs were found in Pb-In NWs [19]. Semi-quantitative studies of $g(E)$ function, which was subject to an undefined prefactor, were reported for Cu films [16] and AuPd NWs [17].

The rutile metal ruthenium dioxide ($RuO_2$) has attracted lasting attention due to its low electrical resistivity [20], thermal and chemical stability under the ambient conditions [4], and high electrocatalytic activities which are useful for oxygen evolution reaction (OER) applications.

In OER applications, such as water splitting [21], oxygen vacancies ($V_O$'s) were found to enhance the electrocatalytic processes [22]. Stoichiometric and nonstoichiometric $RuO_2$ NWs thus have promising potential for a variety of nanotechnological and industrial applications. Recently, we have measured the $g(E)$ functions in a series of thermally annealed $RuO_2$ thin films which contained various amounts of $V_O$'s [23]. Moreover, we have unambiguously detected the spontaneous and repeated motion of nanometer-sized crystallites (hereinafter, called nanocrystallites) in several $RuO_2$ NWs at room temperature [24]. In the latter case, the bonding strength between grain boundaries (GBs) was inferred through an assumption that the nanocrystallite kinetics must be governed by a thermal activation process. In this work we have carried out further studies along these directions with an aim to gaining a deeper understanding of the microscopic characteristics of dynamic atomic and nanocrystalline defects in $RuO_2$ NWs. We focus on several NWs with diameters ranging from 30 to 75 nm. Due to the large surface-to-volume ratios and the relatively small (~10 nm) grain sizes in such small-diameter NWs, we expect to find pronounced electrical noise caused by defect motion. For reference, grain sizes are usually much larger (~50–200 nm) in polycrystalline metal films [14,16]. Thus, GBs shall play comparatively significant roles on generating noise in NWs than in thin films. We first report quantitatively extracted $g(E)$ functions for $RuO_2$ NWs. Second, we demonstrate in a direct manner the thermally activated motion of nanocrystallites around room temperature. The kinetics of these dynamic nanocrystalline defects is then quantitatively analyzed.

## II. EXPERIMENTAL METHOD

**Nanowire grow and device fabrications.** Our $RuO_2$ NWs were grown via metal-organic chemical vapor deposition (MOCVD) on sapphire substrates, with the source reagent bis(ethylcyclopentadienyl)ruthenium [25,26]. High-purity oxygen with a flow rate of 100 sccm was used as both carrier gas and reactive gas. The substrate temperature and pressure of the CVD chamber were controlled at 450°C and 10–50 torr, respectively, during NW growth. After the growth, individual NWs were transferred to a Si substrate capped with a 300-nm-thick $SiO_2$ layer. Metallic Cr/Au (10/100 nm) submicron electrodes were thermally deposited using the standard electron-beam-lithography technique to form contacts with single NWs. The contact resistance between the NW and electrode was typically ~ 100 to 500 Ω at room temperature.

**Electrical noise measurements.** The electrical noise measurement circuit is schematically depicted in Fig. 1(b), which was similar to that utilized in our previous work [24].

The value of the ballast resistor was $R_b = 1$ MΩ, being much larger than the NW resistance (~500 Ω to ~10 kΩ). The measurement principle was based on the modulation and demodulation technique similarly to that described in Refs. [15,27], where the $1/f$ noise contribution from the preamplifier (Stanford Research Systems model SR560) was largely minimized and the measurement sensitivity reached the lowest noise level of the preamplifier. According to the noise contours of SR560, an ac driving current with a carrier frequency $f_c \approx 3$ kHz was applied. The sample resistance at time $t$ was written by $r(t) = \langle r \rangle + \delta r(t)$, where $\langle r \rangle$ was the time-averaged resistance, and $\delta r(t)$ was the temporal resistance fluctuations. Under an ac current with the frequency $f_c$, $\delta r(t)$ was transferred to the temporal voltage fluctuations $\delta V_s(t)$ with frequencies centering at $f_c$. After $\delta V_s(t)$ was amplified by SR560, the signal was phase-sensitively detected by a lock-in amplifier (Stanford Research Systems model SR830), and the frequencies of the noise signal were shifted (demodulated) from $\sim f_c$ back to $f$ to recover the original frequencies of $\delta r(t)$ [15,27]. After the demodulation, the signal was measured by the dynamic signal analyzer (Stanford Research Systems model SR785). The SR785 read the voltage signal with a sampling rate of 1024 Hz and stored the readings in the buffers. A computer fetched the data from the buffers and calculated the voltage noise power spectrum density (PSD) $S_V(f,T)$. The measurements of the noise at various temperature $T$ values were performed on a standard liquid-helium cryostat. The device temperature was monitored with a calibrated silicon diode thermometer. Table 1 lists the relevant parameters for the RuO$_2$ NWs A to E studied in this work. NWs A to D were as-grown, while NW E was focused ion beam (FIB) irradiated.

### III. RESULTS AND DISCUSSION

#### A. Energy distribution function for mobile atomic defects

We have measured the $1/f$ noise between 80 and 370 K to determine the $g(E)$ function and the associated number density of atomic TLSs ($n_{TLS}$) in several polycrystalline RuO$_2$ NWs. Figure 1(b) shows a scanning electron microscopy image for NW A and the electrical circuit for measuring the temporal resistivity fluctuations $\rho(t)$. The $T$ dependence of the time-averaged resistivity $\bar{\rho}$ for our NWs reveals typical metallic behavior, i.e., $\bar{\rho}$ decreases with decreasing $T$ [28,29]. The inset of Fig. 2(a) shows the $\bar{\rho}(T)$ data for NWs A to C. For an Ohmic conductor under a constant bias current, the $\rho(t)$ fluctuations result in a voltage noise PSD $S_V(f,T)$ which is generally expressed by the empirical form [9]: $S_V(f,T) = \gamma\left(V^2/N_c f^\beta\right) + S_V^0$, where $\gamma = \gamma(T)$ is the (dimensionless) Hooge parameter which characterizes the noise magnitude, $N_c$ is the total number of charge carriers in the sample, $V$ is the bias voltage across the sample, $\beta \approx 1$ is the frequency exponent (see below for further discussion), and $S_V^0$ is the background noise stemmed from the thermal noise and the input noise of the preamplifier. Figure 1(c) shows the variation of $S_V(f, T = 300\text{ K})$ with frequency for NW A at three bias voltages. For the lowest (and nominally negligible) bias $V = 0.006$ mV, the $1/f^\beta$ or $1/f$ noise PSD is small, indicating that the measured $S_V(f)$ (at $f > 0.1$ Hz) is dominated by the frequency independent $S_V^0$. As bias $V$ increases to tenths meV, the $1/f$ noise increases and dominates the measured $S_V(f)$ at low frequencies. Figure 1(d) shows the $V^2$ dependence of $\langle S_V \times f \rangle$ for NW A at three $T$ values. Here $\langle S_V \times f \rangle = \gamma V^2/N_c + \langle f \rangle S_V^0$ is the average of the product of each discrete $(f_i, S_{Vi})$ reading in the data set, and $\langle f \rangle$ denotes the average of $f_i$ in the frequency range $0.03 \leq f_i \leq 3$ Hz. At each bias $V$, ten $S_V(f)$ curves have been measured and averaged to calculate $\langle S_V \times f \rangle$. The $\gamma$ values at various temperatures can thus be accurately determined from the slopes $(= \gamma/N_c)$ of the straight dashed lines in Fig. 1(d), without the necessity for subtracting the background contribution $S_V^0$ [23]. This method has the advantage of minimizing any uncertainty which might be introduced to $S_V \times f$ if one only calculates the product from a single frequency value, as was usually done in the literature [5,14,16,17,30,31]. Also, because $RuO_2$ is a normal metal [32], we can safely use a $T$ independent charge carrier number density $n_e \approx 1\times10^{28}$ m$^{-3}$ [33] to calculate the $N_c$ and thus $\gamma(T)$ values for all NWs. We mention that for interconnect applications, quantitative information about $\gamma(T)$ above room temperature is desirable, because the maximum operating $T$ may reach as high as ~100°C, for example, in a computer CPU.

Figure 2(a) shows the variation of the measured $\gamma(T)$ data with $T$ for NWs A to C. The $\gamma$ value remains roughly constant in the temperature range 80–270 K. Between ~270 and 370 K, $\gamma$ increases suddenly with increasing $T$. According to the Dutta-Dimon-Horn (DDH) model [31], which considers an ensemble of thermally activated processes, the frequency exponent $\beta$ in the

$1/f^\beta$ noise is connected to the temperature dependence of $S_V(f,T)$ through the relation

$$\beta(T) = 1 - \frac{1}{\ln(2\pi f \tau_0)}\left[\frac{\partial \ln S_V(T)}{\partial \ln T} - 1\right] = 1 - \frac{1}{\ln(2\pi f \tau_0)}\left[\frac{\partial \ln \gamma(T)}{\partial \ln T} - 1\right] \quad , \quad \text{where} \quad \tau_0 \quad \text{is a}$$

characteristic attempt time. Typically, $\tau_0 \sim 10^{-14}$ s in solids [9]. Thus, the DDH model predicts $\beta < 1$ ($\beta > 1$) for $T < 270$ K ($T > 270$ K) in those RuO$_2$ NWs whose $\gamma(T)$ data are shown in Fig. 2(a). From the $S_V(f,T) \propto f^{-\beta}$ curves measured at various temperatures and at bias voltages $\approx 0.7$–2 mV (where the $1/f^\beta$ noise dominates over $S_V^0$), we obtain average values of $\beta \simeq 0.91$ (1.20), 0.89 (1.23), and 0.89 (1.12) for NWs A, B and C, respectively, at $T$ lower (higher) than ~270 K. These values are in good accord with the prediction of the DDH model, strongly indicating that the observed $1/f^\beta$ noise originate from an ensemble of thermally activated atomic TLSs.

With the $\gamma$ value obtained, the energy distribution function for the atomic TLSs can be determined. Theoretically, $g(\tilde{E})$ at the particular energy $E = \tilde{E}$ is calculated from the $\gamma$ value measured at the corresponding $T$ value which satisfies the relation $\tilde{E}(T) = -k_B T \ln(2\pi f \tau_0)$ [9]. In terms of the electronic parameters and the measured $\rho(T)$ and $\gamma(T)$ values of the given device, $g(\tilde{E})$ can be written by [23,34]

$$g(\tilde{E}) \simeq \frac{4\pi n_e \gamma(T)}{2.6 k_B T}\left[\frac{\rho(T) e^2}{m v_F \langle \sigma_c \rangle}\right]^2 , \quad (1)$$

where $k_B$ is the Boltzmann constant, $e$ is the electronic charge, $m$ is the effective electron mass, $v_F$ ($k_F$) is the Fermi velocity (wavenumber), and $\langle \sigma_c \rangle \approx 4\pi/k_F^2$ is the average scattering cross-section for a moving atomic defect. Thus, through measuring $\gamma(T)$ and $\rho(T)$ over a range of $T$, a finite energy range of the $g(\tilde{E})$ function can be extracted. In RuO$_2$ [32], the electronic parameters are $m \approx 1.4\, m_0$ ($m_0$ being the free electron mass), $v_F \approx 8\times10^5$ m/s, and $k_F \approx 1\times10^{10}$ m$^{-1}$.

Figure 2(b) shows the extracted $g(E)$ functions for NWs A to C, where $E$ was calculated with $f = 1$ Hz. We find that $g(E)$ decreases slightly as $E$ increases from ~0.2 to ~0.6 eV. Then, it rapidly increases at $E > 0.6$ (0.8) eV in NW B (NWs A and C). With our highest measurement $T = 370$ K (which is limited by the cryostat setup), we are able to extract the $g(E)$ function up to $E \approx 1$ eV. The slight increase in $g(E)$ below ~0.6 eV is due to a combination of the absence of $T$ dependence of $\gamma$ at $T < 270$ K [Fig. 2(a)], a decreasing measurement $T$, and a weak $T$ dependence of $\rho(T)$ in Eq. (1). A noise magnitude $\gamma$ independence of temperature in a similar $T$ range was observed in single-wall carbon nanotubes [35]. Note that the fast increase in $g(E)$ at $E > 0.6$ eV is an intrinsic feature of the dynamic defects. It is not due to any extra, unintentional electromigration-induced noise [5], because we had focused our measurements on the $S_V \propto V^2$ regime where the current-driven resistance fluctuations did not occur [9]. Although the $\gamma(T)$ of NW B is close to those of NWs A and C, its $g(E)$ function is higher, especially at $E > 0.6$ eV. This arises from the fact that NW B has a higher $\rho(T)$ than those in the other two NWs, see the inset of Fig. 2(a). We stress that Eq. (1) gives a quantitative expression for $g(E)$ without involving any unknown parameters, while the proportional relation $g(E) \propto 2\pi f \left[ S_V(f,T)/k_B T \right]$ is often used in the literature [5,16,17,30,31,35]. The latter is subject to a proportionality constant, thus it cannot distinguish the $g(E)$ values in samples having a similar $\gamma(T)$ but differing $\rho(T)$. Previous measurements on Ag films reported that the $g(E)$ function attained its peak at $E_p \approx 0.9$ eV [31]. Recent microelectronics studies reported $E_p \approx 0.77$ eV for Cu interconnects [5]. Figure 2(b) indicates that a peak in $g(E)$, if exists, will occur at $E_p > 1$ eV. Since the $E_p$ value somewhat reflects the activation energy for the electromigration effect [30], RuO$_2$ NWs may thus be used for reliable interconnections with high breakdown current densities [36].

Recently, from our systematic studies of RuO$_2$ thin films thermally annealed in oxygen and argon gases, we have identified the atomic TLSs to arise from the $V_O$'s [23]. More precisely, in a given RuO$_2$ sample, we have established that the number density of atomic TLSs, $n_{TLS}$, (approximately) equals the concentration of oxygen vacancies, $n_{V_O}$. We have also observed that the thermal annealing effects on RuO$_2$ NWs are similar to those on thin films, i.e., $n_{V_O}$ can be reversibly increased (reduced) by annealing in vacuum or argon (oxygenation in air) [29]. Theoretically, the activation energy $E$ of $V_O$'s on the (110) surface of a RuO$_2$ crystal has been evaluated from the density functional calculations [37]. The results indicate a position dependent $E$ value: $E \sim 0.7$ eV for an on-top O to hop into an adjacent vacancy in the bridging O row, and $E \sim 0.9$ eV for a $V_O$ in the bridging O row to migrate along the row. Intuitively, one would expect a higher $E$ value for $V_O$'s in the bulk than on the surface, due to a larger coordination number of Ru atoms bonding with an O atom. Thus, the observation of $E_p > 1$ eV in Fig. 2(b) is fairly reasonable. Because our NWs are polycrystalline with relatively small grain sizes as mentioned, a certain degree of lattice disorder must exist near GBs, thus leading to a (wide) distribution of $E$ values [23,31]. This can plausibly explain the observation of a small but finite $g(E)$ at $E < 0.6$ eV. Furthermore, although the NWs studied in this work were all taken from the same batch, probably due to their differing positions located on the sapphire substrate during the growth process, the $n_{V_O}$ values may vary among NWs [29]. Thus, $g(E)$ will vary from NW to NW, as found in Fig. 2(b). Numerically, the value of $n_{TLS} \approx n_{V_O}$ can be calculated from $g(E)$ through the equation $n_{TLS}(T) \approx g(E) \times 2.6 k_B T$ [23]. At $T = 300$ K, we obtain $n_{TLS} \approx 2 \times 10^{26}$, $\approx 8 \times 10^{26}$ and $\approx 2 \times 10^{26}$ m$^{-3}$, corresponding to the ratios $n_{V_O} / n_O \approx 0.3\%$, $\approx 1\%$ and $\approx 0.3\%$, in NWs A, B and C, respectively, where $n_O$ is the number density of oxygen atoms in the stoichiometric RuO$_2$ rutile unit cell. This method for inferring the $n_{V_O}$ values is efficient and noninvasive.

### B. Thermally activated nanocrystalline defect motion

High-resolution transmission electron microscopy (HR-TEM) studies have shown that our polycrystalline RuO$_2$ NWs are constituted of nm-sized crystallites [24]. Therefore, GBs exist in a given RuO$_2$ NW. The bonding strength between GB atoms is much weaker than that of bulk atoms residing inside a crystallite. Due to the weak GB bonding strength, one expects that a nanocrystallite may repeatedly switch between two (or several) configurational states. Conceptually, such granular switches will take place occasionally, in contrast to the atomic defect fluctuations that are ubiquitous. To search for mobile nanocrystallites and, in particular, to study their kinetics, we have examined in this and previous works more than 30 as-grown RuO$_2$ NWs, and found RTN signals in five of them (NW D to be discussed below and the four NWs reported in Ref. [24]). Note, however, that the possible existence of dynamic nanocrystallites in other NWs cannot be totally ruled out, because their characteristic switching rates may be beyond our experimental bandwidth (~ 0.03–1000 Hz). Figure 3(a) shows the time-resolved $\rho(t)$ for NW D at 300 K, whose values were registered over 30 s with one reading per 1 ms. Clearly, $\rho$ switches between two resistivity values which differ by $\Delta \rho_{NC} = \rho_1 - \rho_2 \approx 2.6$ $\mu\Omega$ cm, corresponding to $\Delta \rho_{NC} / \bar{\rho} \approx 0.6\%$. This value of $\Delta \rho_{NC}$ is 5 orders of magnitude larger than the resistivity variation ($\Delta\rho_a \approx 1.2 \times 10^{-5}$ $\mu\Omega$ cm) that would arise from the conduction electron scattering off a single atomic TLS [24], strongly suggesting that the observed $\Delta\rho_{NC}$ originate from the electron scattering off a granular TLS. Numerically, the size of the dynamic nanocrystallite may be evaluated through the approximate equation $\Delta\rho_{NC} \approx N_{NC}\Delta\rho_a$, where $N_{NC}$ is the total number of atoms constituting the mobile nanocrystallite [24]. We obtain $N_{NC} \approx 2.1 \times 10^5$ atoms in this case, corresponding to a linear crystallite size of ~16 nm, in line with the HR-TEM results.

The unequal dwell time of the two resistivity states exhibited in Fig. 3(a) points to an asymmetric double-well potential in the TLS model, *cf.* Fig. 1(a). We denote the two states as state 1 (state 2), with barrier height $E_1$ ($E_2$), and the relaxation time from state 1 to state 2 (state 2 to state 1) as $\tau_1$ ($\tau_2$). Assume that $\tau_1 = \alpha\tau_2$, with $\alpha$ being a dimensionless constant. Then, the resistivity noise PSD due to $\rho(t)$ can be expressed by [38,39]

$$S_\rho(f) = 4\alpha \left(\frac{1}{1+\alpha}\right)^2 \left[\frac{(\Delta\rho_{NC})^2 \tau}{1+(2\pi f)^2 \tau^2}\right] + \frac{B}{f} + S_\rho^0, \quad (2)$$

where $1/\tau = 1/\tau_1 + 1/\tau_2$, $B$ is a constant, and $S_\rho^0$ is the background PSD. The first term on the right hand side of Eq. (2), which possesses a Lorentzian form, originates from the bi-state resistivity fluctuations due to a single dynamic nanocrystallite. The second term is the $1/f$ noise stemming from a collection of mobile atomic defects that always exist in a device. Figure 3(b) shows the PSD $S_\rho(f)$ converted from the $\rho(t)$ data in Fig. 3(a). The $f$ dependence of $S_\rho$ can be well described by Eq. (2) (dashed curve), with the fitted values $\tau_1 = 60$ ms, $\tau_2 = 310$ ms, $B = 9.0 \times 10^{-2}$ $\mu\Omega^2$ cm$^2$, and $S_\rho^0 = 2.9 \times 10^{-5}$ $\mu\Omega^2$ cm$^2$ Hz$^{-1}$. We should point out that the Lorentzian term dominates the overall frequency behavior. For example, at $f = 3-5$ Hz, $\approx 75\%$ of the measured $S_\rho$ arises from this term. Figure 3(b) also reveals that the $S_\rho(f)$ curves measured with two different bias currents overlap, suggesting that the observed nanocrystallite motion is intrinsic but not driven by the small bias currents. For a rough estimate, by substituting $\tau \sim 0.1$ s, $\tau_0 \sim 10^{-14}$ s, and $T = 300$ K into Eq. (3), we obtain an activation energy $\bar{E} \sim 0.8$ eV for this nanocrystalline defect motion (see below for further discussion). [Unfortunately, NW D was incidentally burned out while we tried to measure $\rho(t)$ at different temperatures.]

It is a long-standing conjecture that nanocrystalline defect motion (around room $T$) must take place via thermal activation processes. However, to our knowledge, this intuitive notion has never been tested by experiment, partly because mobile nanocrystallites are hard to detect, as mentioned. In practice, studies of the kinetics of granular TLSs can provide valuable information on the important material parameter such as the GB bonding strength. To achieve this aim, it is desirable to (controllably, if at all possible) produce mobile nanocrystallites with their characteristic relaxation time scale(s) lying within the experimental bandwidth. Previously, it has been established that high-energy electron and $Kr^+$-ion irradiation on Cu thin films generated extra mobile defects and enhanced the $1/f$ noise in as-deposited samples [16]. On the contrary, it was found that thermal annealing of AuPd NWs reduced the number densities of mobile defects and thus the $1/f$ noise magnitudes [17]. Similarly, the irradiation-induced mobile defects in Cu films could be annealed away [16]. In those cases, the authors were largely concentrated on the generation and removal of mobile atomic defects. By the same token, the FIB irradiation (with, e.g., $Ga^+$ ions) can also produce additional defects in a given sample [40]. We have therefore applied a FIB to bombard NW E with the hope of producing measurable mobile nanocrystallites, in addition to moving atomic defects (see Table 1 for the FIB irradiation conditions). After the FIB bombardment, NW E has a relatively high resistivity compared with the typical value $\bar{\rho}(300 \text{ K}) \simeq 150-800$ $\mu\Omega$ cm for as-grown $RuO_2$ NWs [28,29,41]. Moreover, its $\bar{\rho}(T)$ curve shows a very weak temperature dependence between 140 and 300 K, with $\bar{\rho}(140 \text{ K}) \simeq 1412$ and $\bar{\rho}(300 \text{ K}) \simeq 1400$ $\mu\Omega$ cm, indicating that the FIB bombardment has introduced a high level of lattice disorder. [We have carried out FIB bombardment on six $RuO_2$ NWs, and observed mobile nanocrystallites in two of them, including NW E. The RTN patterns in the other NW were unreproducible, probably due to an incidentally large bias voltage applied to measure $\rho(t)$. Therefore, we only analyze the data for NW E. These preliminary results suggest that FIB irradiation can provide an effective way for meaningful studies of nanocrystalline defect kinetics.]

Figures 4(a)–(c) show the temporal $\rho(t)$ fluctuations of NW E at three $T$ values. Similar to that in NW D, $\rho(t)$ switches between two values, with $\Delta\rho_{NC} \approx 5-6$ $\mu\Omega$ cm. This resistivity variation corresponds to a significant change in the NW resistivity of $\Delta\rho_{NC}/\bar{\rho} \approx 0.4\%$, suggesting a linear size of ~17 nm for the detected mobile nanocrystallite. The observed unequal dwell times of the two states imply an asymmetric double-well potential of the granular TLS, similar to the case of NW D. Figures 4(d), (e) and (f) show plots of the corresponding histograms

for Figs. 4(a), (b) and (c), respectively. The two peaks in the histogram at each $T$ value can be described by a sum of two Gaussian curves (red curves). Figures 4(a)–(c) demonstrate that, as $T$ increases from 275 to 295 K, the resistivity switching occurs considerably faster, implying that the relaxation times $\tau_1$ and $\tau_2$ are strong functions of $T$.

Figure 5(a) shows the $S_\rho(f)$ curves for NW E at several $T$ values. We see that $S_\rho$ first increases rapidly with decreasing $f$ below ~100 Hz, then the dependence slows down at a crossover frequency $f'$, as indicated by the arrows. Note that $f'$, which is defined by $f' = 1/(2\pi\tau)$ from the first term in Eq. (2), increases with increasing $T$. Thus, $1/\tau$ increases with increasing $T$. By fitting the $S_\rho(f,T)$ curve at each temperature to Eq. (2) (dashed curves), we obtain the values for relaxation times $\tau(T)$, $\tau_1(T)$ and $\tau_2(T)$, and the parameters $B = (2.4-8.8)\times 10^{-2}$ $\mu\Omega^2$ cm$^2$, and $S_\rho^0 = (0.83-1.2)\times 10^{-3} \mu\Omega^2$ cm$^2$ Hz$^{-1}$. Figure 5(b) shows the $T$ dependence of these relaxation times between 270 and 305 K. A $\ln(\tau_i) \propto T^{-1}$ law ($i$ = 1, 2) is observed, unambiguously demonstrating that the granular TLS fluctuations are governed by the thermal activation processes. The relaxation rates, *cf.* Fig. 1(a), are given by

$$\frac{1}{\tau_i} = f_{0,i} \exp\left(\frac{-E_i}{k_B T}\right), \quad (3)$$

where $f_{0,i}$ denotes the attempt frequency at state $i$. From linear fits to Eq. (3) (straight dashed lines), we obtain $E_1 \approx 0.85$ eV and $E_2 \approx 0.68$ eV. The asymmetry of the TLS is $\Delta E = E_1 - E_2 \approx 0.17$ eV, or $\Delta E/\overline{E} \approx 22\%$, where $\overline{E} = (E_1+E_2)/2$. This average barrier height $\overline{E} \approx 0.77$ eV is in agreement with that estimated for NW D and with several previously studied RuO$_2$ NWs [24], where the $\rho(t)$ fluctuations were measured only at $T$ = 300 K and the $\overline{E}$ values were inferred from a less direct method. Note that this $\overline{E}$ value is more than one order of magnitude smaller than the interatomic bonding strength ($\sim 10-30$ eV) for bulk atoms in a crystallite [42]. This is in agreement with the reverse Hall-Petch effect which states that the mechanical strength decreases in a material composed of nm-sized grains [43,44].

The above approach for extracting $\overline{E}$ value through RTN measurements is relevant to a variety of nanotechnological applications. For example, due to the weakened bonding strength, GB atoms may be driven to move under a high current density, causing electromigration in nanoelectronic devices. It was recently found that the GB stability in nanograined Ni-Mo metal was enhanced by thermal annealing, resulting in increased material hardness [45]. In 2D materials, an increased $1/f$ noise in the presence of GBs was observed [46]. In addition, it was reported that GBs in a 2D material could be used to fabricate memristors [47]. Characterizing GB properties through RTN studies, as shown in this work, will help to gain insight into these issues. Because RTN switching causes much larger resistivity fluctuations than $1/f$ noise does, its pernicious effect on nanodevices could be even worse than the mobile atomic defects. This issue has not been much addressed in the literature while it should deserve close studies.

## IV. CONCLUSION

As downsizing of nanodevice dimensions continues, undesirable dynamic defects, which exist ubiquitously in real materials and cause low-frequency noise and structural instability, play detrimental roles on limiting the ultimate performance and reliability of miniaturized devices. The $RuO_2$ rutile has witnessed a wide range of technological and industrial applications for decades. Unlike most nanostructures made of other materials, $RuO_2$ NWs are chemically and thermally stable in the ambient. Moreover, its oxygen contents can be reversibly adjusted via, for example, thermal annealing or ionic-liquid gating [48]. Very recently, theoretical and experimental studies have discovered that $RuO_2$ is a Dirac nodal line metal, making it a fascinating member of the rutile family [49]. Through investigations of electronic noise processes, we have extracted the activation energy distribution function $g(E)$ which characterizes the energy and time scales of moving atomic defects in polycrystalline $RuO_2$ NWs. In addition, we have demonstrated that around room temperature nanocrystalline defect motion clearly exists, which is driven by thermal activation. Our quantitative results provide prudent information about differing types of dynamic defects in metal nanostructures.

## ACKNOWLEDGMENTS

The authors are grateful to the late Y. S. Huang for growing $RuO_2$ NWs. This work was supported by Ministry of Science and Technology, Taiwan (grant No. MOST 106-2112-M-009-007-MY4) and the Center for Emergent Functional Matter Science of National Chiao Tung University from The Featured Areas Research Center Program within the framework of the Higher Education Sprout Project by the Ministry of Education (MOE) in Taiwan.

# REFERENCES

[1] J. Moser, J. Güttinger, A. Eichler, M. J. Esplandiu, D. E. Liu, M. I. Dykman, and A. Bachtold, Ultrasensitive force detection with a nanotube mechanical resonator, Nat. Nanotechnol. **8**, 493 (2013).

[2] G. Zolfagharkhani, A. Gaidarzhy, P. Degiovanni, S. Kettemann, P. Fulde, and P. Mohanty, Nanomechanical detection of itinerant electron spin flip, Nat. Nanotechnol. **3**, 720 (2008).

[3] A. K. Naik, M. S. Hanay, W. K. Hiebert, X. L. Feng, and M. L. Roukes, Towards single-molecule nanomechanical mass spectrometry, Nat. Nanotechnol. **4**, 445 (2009).

[4] J. V. Ryan, A. D. Berry, M. L. Anderson, J. W. Long, R. M. Stroud, V. M. Cepak, V. M. Browning, D. R. Rolison, and C. I. Merzbacher, Electronic connection to the interior of a mesoporous insulator with nanowires of crystalline $RuO_2$, Nature **406**, 169 (2000).

[5] G. Liu, S. Rumyantsev, M. A. Bloodgood, T. T. Salguero, M. Shur, and A. A. Balandin, Low-Frequency Electronic Noise in Quasi-1D $TaSe_3$ van der Waals Nanowires, Nano Lett. **17**, 377 (2017).

[6] T. A. Empante *et al.*, Low Resistivity and High Breakdown Current Density of 10 nm Diameter van der Waals $TaSe_3$ Nanowires by Chemical Vapor Deposition, Nano Lett. **19**, 4355 (2019).

[7] M. Imboden and P. Mohanty, Dissipation in nanoelectromechanical systems, Phys. Rep. **534**, 89 (2014).

[8] A. N. Cleland and M. L. Roukes, Noise processes in nanomechanical resonators, J. Appl. Phys. **92**, 2758 (2002).

[9] P. Dutta and P. M. Horn, Low-frequency fluctuations in solids: $1/f$ noise, Rev. Mod. Phys. **53**, 497 (1981).

[10] A. A. Balandin, Low-frequency $1/f$ noise in graphene devices, Nat. Nanotechnol. **8**, 549 (2013).

[11] P. W. Anderson, B. I. Halperin, and C. M. Varma, Anomalous low-temperature thermal properties of glasses and spin glasses, Philos. Mag. **25**, 1 (1972).

[12] W. A. Phillips, Tunneling states in amorphous solids, J. Low Temp. Phys. **7**, 351 (1972).

[13] R. D. Black, P. J. Restle, and M. B. Weissman, Nearly Traceless $1/f$ Noise in Bismuth, Phys. Rev. Lett. **51**, 1476 (1983).

[14] R. H. Koch, J. R. Lloyd, and J. Cronin, $1/f$ Noise and Grain-Boundary Diffusion in Aluminum and Aluminum Alloys, Phys. Rev. Lett. **55**, 2487 (1985).

[15] J. Pelz and J. Clarke, Dependence of $1/f$ Noise on Defects Induced in Copper Films by Electron Irradiation, Phys. Rev. Lett. **55**, 738 (1985).

[16] J. Pelz, J. Clarke, and W. E. King, Flicker ($1/f$) noise in copper films due to radiation-induced defects, Phys. Rev. B **38**, 10371 (1988).

[17] D. M. Fleetwood and N. Giordano, Direct link between $1/f$ noise and defects in metal films, Phys. Rev. B **31**, 1157 (1985).

[18] K. S. Ralls and R. A. Buhrman, Defect Interactions and Noise in Metallic Nanoconstrictions, Phys. Rev. Lett. **60**, 2434 (1988).

[19] N. Giordano and E. R. Schuler, Giant conductance fluctuations in thin metal wires, Phys. Rev. B **41**, 11822 (1990).

[20] W. D. Ryden, A. W. Lawson, and C. C. Sartain, Electrical Transport Properties of $IrO_2$ and $RuO_2$, Phys. Rev. B **1**, 1494 (1970).

[21] H. Liu, G. Xia, R. Zhang, P. Jiang, J. Chen, and Q. Chen, MOF-derived $RuO_2/Co_3O_4$ heterojunctions as highly efficient bifunctional electrocatalysts for HER and OER in alkaline solutions, RSC Adv. **7**, 3686 (2017).

[22] Y. Tian *et al.*, A Co-Doped Nanorod-like $RuO_2$ Electrocatalyst with Abundant Oxygen Vacancies for Acidic Water Oxidation, iScience **23**, 100756 (2020).

[23] S.-S. Yeh, K. H. Gao, T.-L. Wu, T.-K. Su, and J.-J. Lin, Activation Energy Distribution of Dynamical Structural Defects in $RuO_2$ Films, Phys. Rev. Appl. **10**, 034004 (2018).

[24] S.-S. Yeh, W.-Y. Chang, and J.-J. Lin, Probing nanocrystalline grain dynamics in nanodevices, Sci. Adv. **3**, e1700135 (2017).

[25] C. C. Chen, R. S. Chen, T. Y. Tsai, Y. S. Huang, D. S. Tsai, and K. K. Tiong, The growth and characterization of well aligned $RuO_2$ nanorods on sapphire substrates, J. Phys. Condens. Matter **16**, 8475 (2004).

[26] Y. H. Lin, Y. C. Sun, W. B. Jian, H. M. Chang, Y. S. Huang, and J. J. Lin, Electrical transport studies of individual $IrO_2$ nanorods and their nanorod contacts, Nanotechnology **19**, 045711 (2008).

[27] J. H. Scofield, AC method for measuring low-frequency resistance fluctuation spectra, Rev. Sci. Instrum. **58**, 985 (1987).

[28] Y.-H. Lin, S.-P. Chiu, and J.-J. Lin, Thermal fluctuation-induced tunneling conduction through metal nanowire contacts, Nanotechnology **19**, 365201 (2008).

[29] S.-S. Yeh, T.-K. Su, A.-S. Lien, F. Zamani, J. Kroha, C.-C. Liao, S. Kirchner, and J.-J. Lin, Vacancy-driven multichannel Kondo physics in topological quantum material $IrO_2$, arXiv:1910.13648 (2019).

[30] D. M. Fleetwood *et al.*, Low-frequency noise and defects in copper and ruthenium resistors, Appl. Phys. Lett. **114**, 203501 (2019).

[31] P. Dutta, P. Dimon, and P. M. Horn, Energy Scales for Noise Processes in Metals, Phys. Rev. Lett. **43**, 646 (1979).

[32] K. M. Glassford and J. R. Chelikowsky, Electronic and structural properties of $RuO_2$, Phys. Rev. B **47**, 1732 (1993).

[33] W.-C. Shin and S. G. Yoon, Characterization of $RuO_2$ Thin Films Prepared by Hot-Wall Metallorganic Chemical Vapor Deposition, J. Electrochem. Soc. **144**, 1055 (1997).

[34] J. Pelz and J. Clarke, Quantitative "local-interference" model for $1/f$ noise in metal films, Phys. Rev. B **36**, 4479 (1987).

[35] D. Tobias, M. Ishigami, A. Tselev, P. Barbara, E. D. Williams, C. J. Lobb, and M. S. Fuhrer, Origins of $1/f$ noise in individual semiconducting carbon nanotube field-effect transistors, Phys. Rev. B **77**, 033407 (2008).

[36] Y. Lee, B.-U. Ye, H. k. Yu, J.-L. Lee, M. H. Kim, and J. M. Baik, Facile Synthesis of Single Crystalline Metallic $RuO_2$ Nanowires and Electromigration-Induced Transport Properties, J. Phys. Chem. C **115**, 4611 (2011).

[37] A. P. Seitsonen and H. Over, in *High Performance Computing in Science and Engineering, Munieh 2002*, edited by S. Wagner *et al.* (Springer-Verlag, Berlin, 2003).

[38] S. Kogan, *Electronic Noise and Fluctuations in Solids* (Cambridge University Press, Cambridge, 1996).

[39] V. K. Sangwan, H. N. Arnold, D. Jariwala, T. J. Marks, L. J. Lauhon, and M. C. Hersam, Low-Frequency Electronic Noise in Single-Layer $MoS_2$ Transistors, Nano Lett. **13**, 4351 (2013).

[40] F. Hofmann *et al.*, 3D lattice distortions and defect structures in ion-implanted nano-crystals, Sci. Rep. **7**, 45993 (2017).

[41] A.-S. Lien, L. Y. Wang, C. S. Chu, and J.-J. Lin, Temporal universal conductance fluctuations in $RuO_2$ nanowires due to mobile defects, Phys. Rev. B **84**, 155432 (2011).

[42] Y. M. Galperin, V. G. Karpov, and V. I. Kozub, Localized states in glasses, Adv. Phys. **38**, 669 (1989).

[43] J. Schiøtz, F. D. Di Tolla, and K. W. Jacobsen, Softening of nanocrystalline metals at very small grain sizes, Nature **391**, 561 (1998).

[44] H. Van Swygenhoven, Grain Boundaries and Dislocations, Science **296**, 66 (2002).

[45] J. Hu, Y. N. Shi, X. Sauvage, G. Sha, and K. Lu, Grain boundary stability governs hardening and softening in extremely fine nanograined metals, Science **355**, 1292 (2017).

[46] V. Kochat, C. S. Tiwary, T. Biswas, G. Ramalingam, K. Hsieh, K. Chattopadhyay, S. Raghavan, M. Jain, and A. Ghosh, Magnitude and Origin of Electrical Noise at Individual Grain Boundaries in Graphene, Nano Lett. **16**, 562 (2016).

[47] V. K. Sangwan, D. Jariwala, I. S. Kim, K.-S. Chen, T. J. Marks, L. J. Lauhon, and M. C. Hersam, Gate-tunable memristive phenomena mediated by grain boundaries in single-layer $MoS_2$, Nat. Nanotechnol. **10**, 403 (2015).

[48] J. Jeong, N. Aetukuri, T. Graf, T. D. Schladt, M. G. Samant, and S. S. P. Parkin, Suppression of Metal-Insulator Transition in $VO_2$ by Electric Field–Induced Oxygen Vacancy Formation, Science **339**, 1402 (2013).

[49] V. Jovic *et al.*, Dirac nodal lines and flat-band surface state in the functional oxide $RuO_2$, Phys. Rev. B **98**, 241101 (2018).

**FIGURES**

**Figure 1**

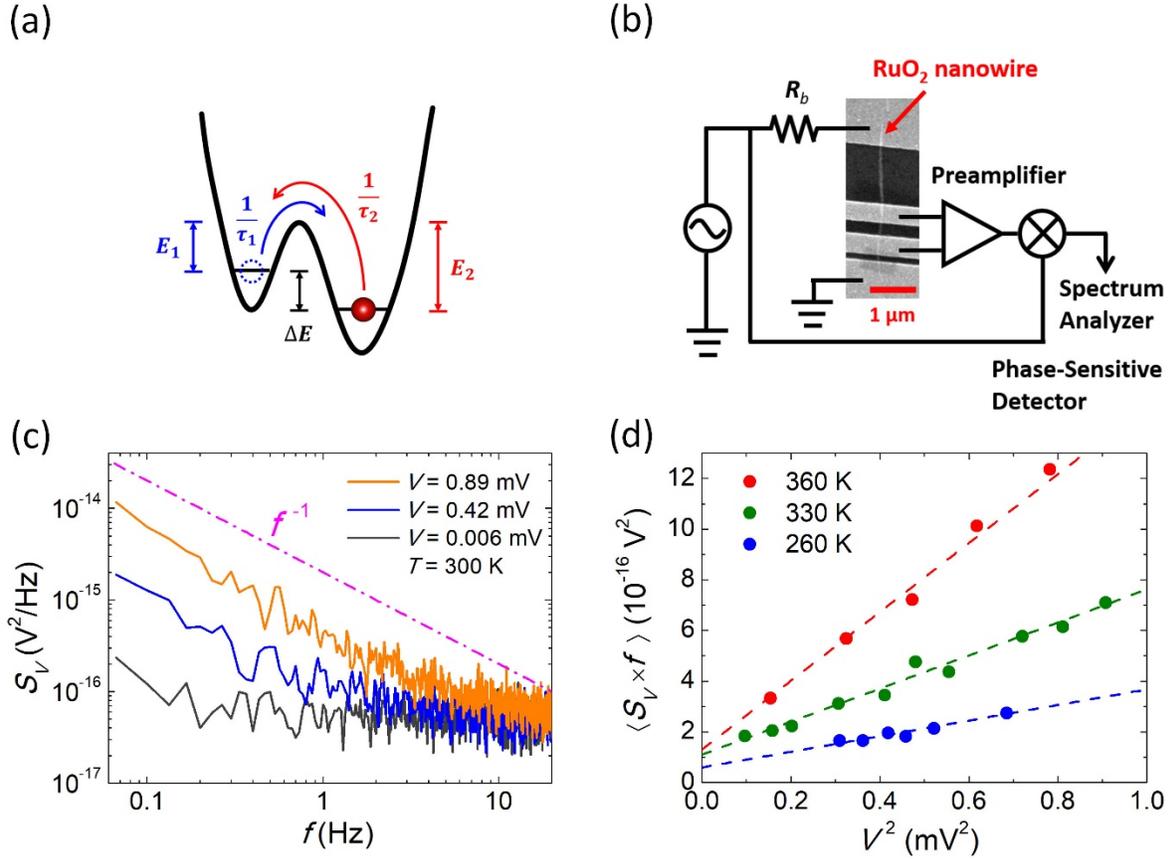

**Figure 1. Two-level system (TLS) model and low-frequency noise measurements.** (a) A schematic double-well potential in the TLS model. The fluctuating object can be a single atom or a cluster of atoms (nanocrystallite in this study). $E_i$ ($i$ = 1, 2) is the barrier height, $1/\tau_i$ is the relaxation rate, and the asymmetry $\Delta E = |E_1 - E_2|$. (b) A schematic diagram depicting the electrical circuit for temporal resistivity fluctuation measurements. $R_b$ is a ballast resistor. Shown in the scanning electron microscopy image is the NW A. (c) Voltage noise power spectrum density $S_V (300 \text{ K})$ as a function of frequency for NW A at three bias voltages ($V$). The dash-dotted line indicates $f^{-1}$ frequency dependence, and is a guide to the eye. (d) $\langle S_V \times f \rangle$ as a function of $V^2$ for NW A at three $T$ values, where $\langle ... \rangle$ denotes average (see text).

**Figure 2**

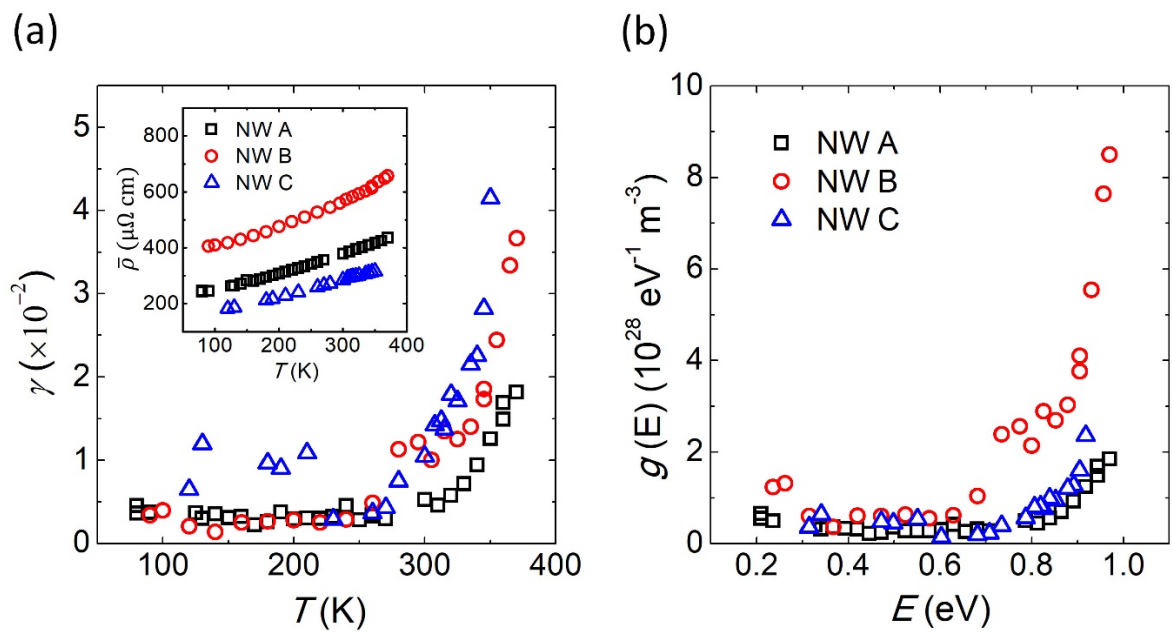

**Figure 2. Low-frequency noise magnitude and energy distribution function.** (a) $1/f$ noise magnitude $\gamma$ (Hooge parameter) as a function of temperature for NWs A, B and C. Inset: time-averaged resistivity $\bar{\rho}$ as a function of temperature for NWs A to C, as indicated. (b) Variation of activation energy distribution function $g(E)$ with energy for NWs A to C. The $g(E)$ values are calculated from the corresponding $\gamma$ values shown in (a).

**Figure 3**

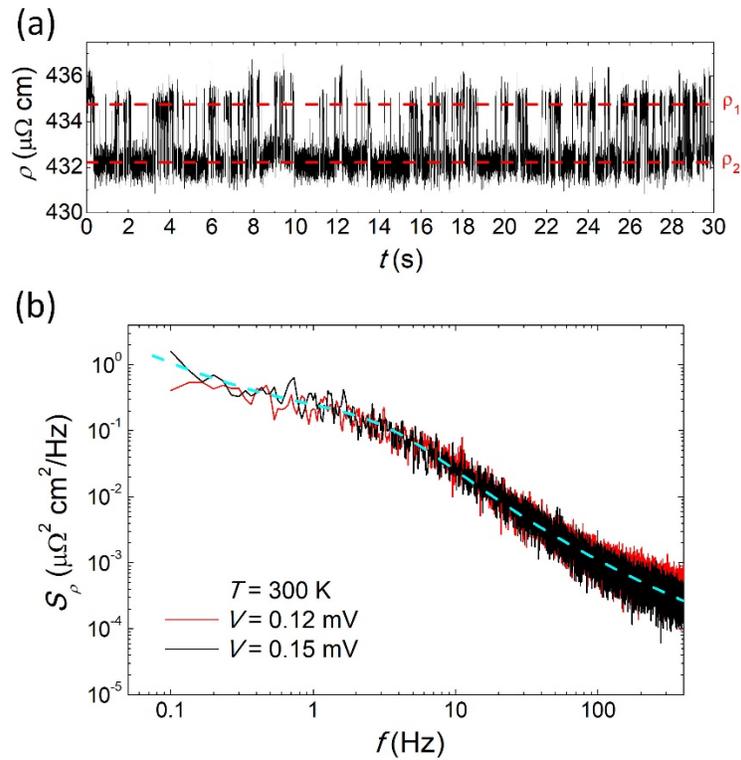

**Figure 3. Temporal resistivity fluctuations and noise power spectrum density (PSD).** (a) Temporal resistivity fluctuations $\rho(t)$ for NW D at 300 K measured with a bias voltage $V = 0.15$ mV. Random telegraph noise (RTN) behavior with bi-state values $\rho_1$ and $\rho_2$ is evident. (b) Resistivity noise PSD $S_\rho$ as a function of frequency for the same NW measured at two bias voltages, as indicated. The dashed curve is a least-squares fit to Eq. (2) (see text).

**Figure 4**

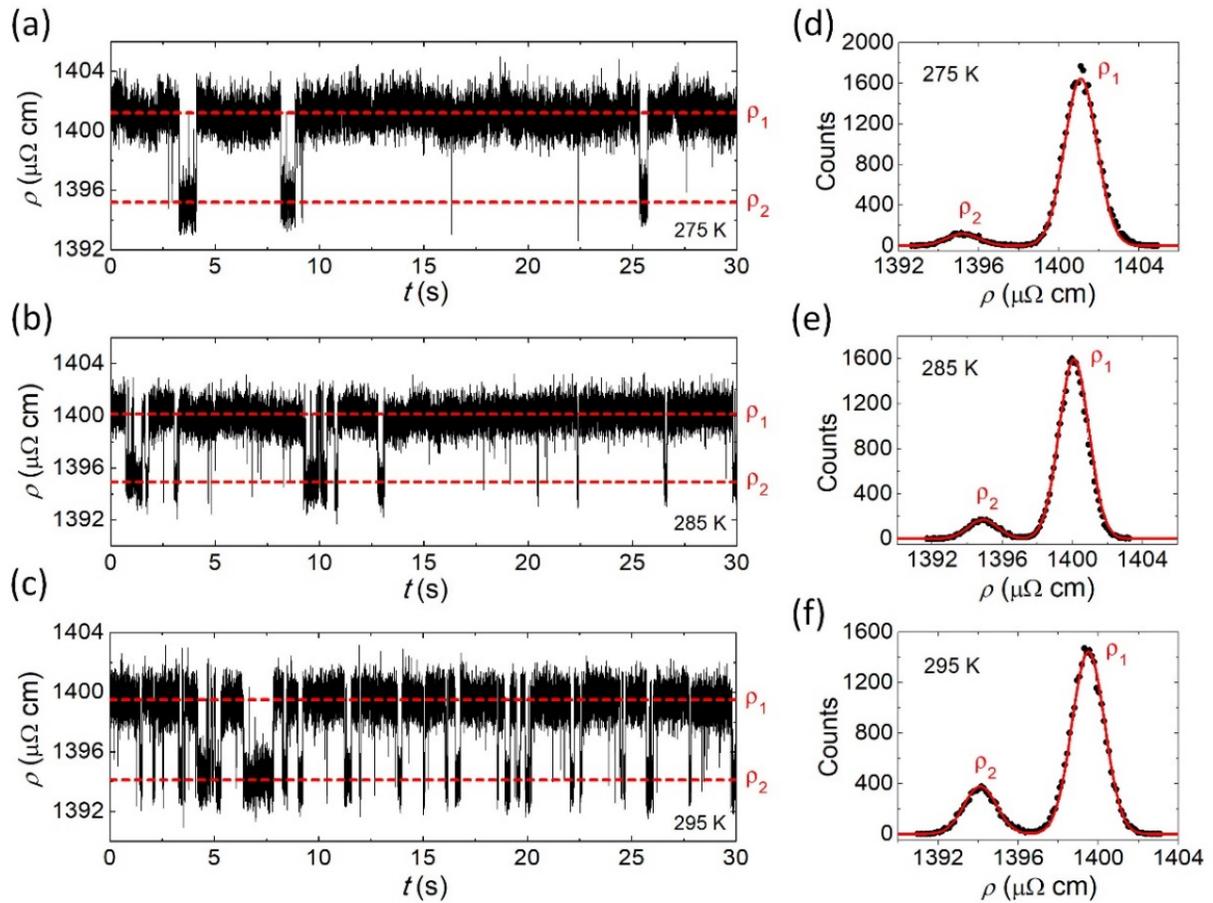

**Figure 4. Random telegraph noise and histograms.** Temporal resistivity fluctuations $\rho(t)$ for NW E at (a) 275, (b) 285, and (c) 295 K, measured with a bias voltage $V \simeq 0.70$ mV. Random telegraph noise (RTN) behavior with bi-state values $\rho_1$ and $\rho_2$ is evident. (d)–(f) Histograms for the corresponding bi-state resistivities shown in (a)–(c). The red solid curve in each panel is a sum of two Gaussian curves, and is a guide to the eye.

**Figure 5**

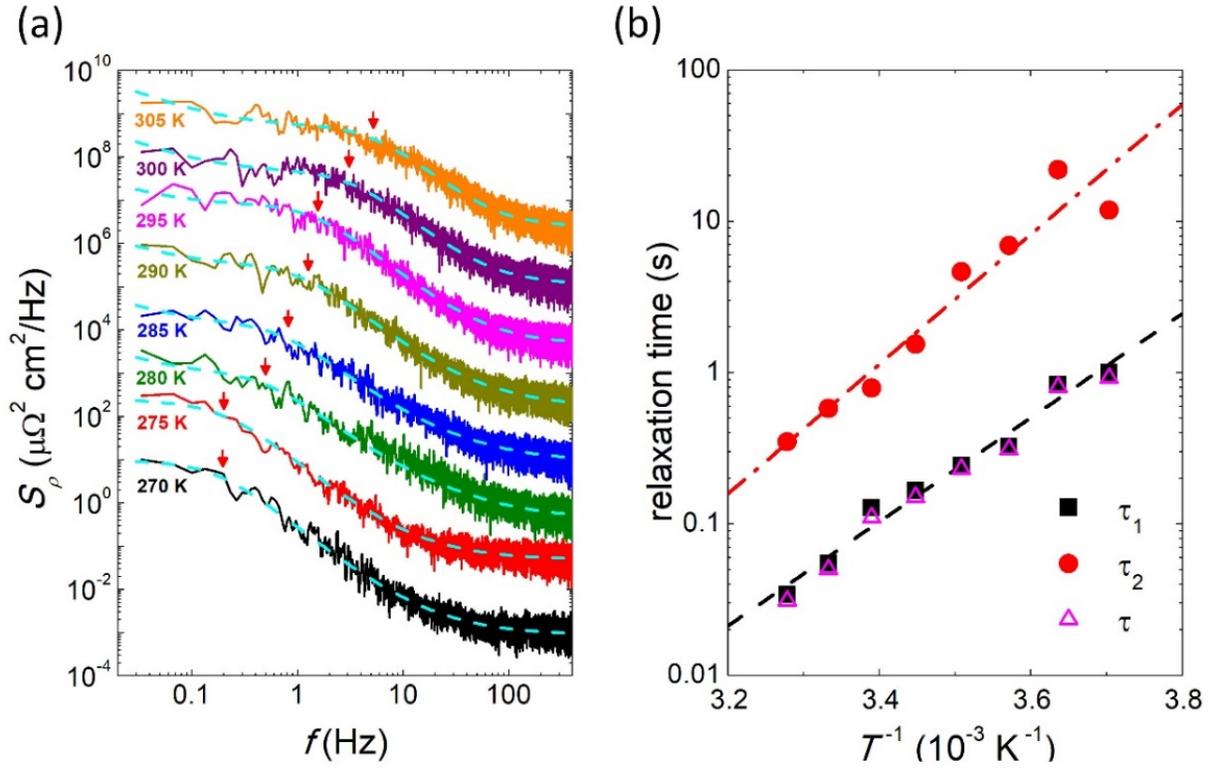

**Figure 5. Resistivity noise power spectrum density and nanocrystallite relaxation times.** (a) $S_\rho$ as a function of frequency for NW E at several temperatures. The dashed curves are least-squares fits to Eq. (2), as described in the text. The arrows indicate the crossover frequency given by $f' = 1/(2\pi\tau)$, where $1/\tau$ is the effective relaxation rate for the fluctuating nanocrystallite. For clarity, the curves from bottom to top are multiplied by factors of 1, 50, 500, $10^4$, $2\times10^5$, $5\times10^6$, $10^8$ and $2\times10^9$, respectively. (b) Relaxation times $\tau_1$, $\tau_2$ and $\tau$ (in a logarithmic scale) as a function of the inverse of temperature. The black and red straight dashed lines are linear fits to $\tau_1$ and $\tau_2$, respectively.

TABLE

**Table 1. Relevant parameters for RuO₂ NWs.** $L$ is the NW length between the voltage probes in a 4-probe configuration, $d$ is the NW diameter, $\rho_{RT}$ is the NW resistivity at room temperature. $T_{LH}$ denotes the temperature range for the noise measurement. $1/f$ and RTN stand for the noise form revealed by the measured power spectrum density (PSD) for each NW. NWs A to D were as-grown. The 4-probe NW E device was irradiated by a FIB (TESCAN GAIA3 FIB-SEM) to produce additional lattice defects. The FIB imaging was performed at 2000× magnification, by employing Ga⁺ ions with an accelerating voltage of 30 keV and an emission current of 10 pA for ~5 s. The NW diameter was measured after the FIB irradiation.

| Nanowires | $L$ (μm) | $d$ (nm) | $\rho_{RT}$ (μΩ cm) | $T_{LH}$ (K) | Noise PSD |
|---|---|---|---|---|---|
| NW A | 0.69 | 40 | 380 | 80–370 | $1/f$ |
| NW B | 0.72 | 30 | 570 | 80–370 | $1/f$ |
| NW C | 1.7 | 35 | 280 | 80–370 | $1/f$ |
| NW D | 0.75 | 75 | 430 | 300 | RTN |
| NW E | 1.33 | 40 | 1400 | 270–305 | RTN |